\documentclass[onecolumn]{aa} 
\usepackage{graphicx}
\usepackage{txfonts}
\begin{document}

\title{Statistics of the two-point cross-covariance function of solar oscillations}


\author{Kaori Nagashima\inst{1} \and Takashi Sekii\inst{2,3}
 \and Laurent Gizon\inst{1,4} \and Aaron C. Birch\inst{1} }

\institute{Max-Planck-Institut f\"ur Sonnensystemforschung, 
 Justus-von-Liebig-Weg 3, 37077 G\"ottingen, Germany\\
              \email{nagashima@mps.mpg.de}
         \and
National Astronomical Observatory of Japan, Mitaka, Tokyo, 181-8588 Japan 
          \and
Department of Astronomical Science, SOKENDAI (the Graduate University for Advanced Studies), Mitaka, Tokyo, 181-8588 Japan
 \and
Institut f\"ur Astrophysik, Georg-August-Universit\"at G\"ottingen, 
Friedrich-Hund-Platz 1, 37077  G\"ottingen, Germany
 }

   \date{Received January 14, 2016; accepted June 17, 2016}

 
\abstract
{The cross-covariance of solar oscillations observed at pairs of points on the solar surface is a fundamental ingredient in time-distance helioseismology. Wave travel times are extracted from the cross-covariance function and are used to infer the physical conditions in the solar interior.
}{  
Understanding the statistics of the two-point cross-covariance function is a necessary step towards optimizing the measurement of travel times.
}{
By modeling stochastic solar oscillations, we evaluate the variance of the cross-covariance function as function of time-lag and distance between the two points. 
}{
We show that the variance of the cross-covariance is independent of both time-lag and distance in the far field, i.e., when they are large compared to the coherence scales of the solar oscillations.
}{The constant noise level for the cross-covariance means that the signal-to-noise ratio for the cross-covariance is proportional to the amplitude of the expectation value of the cross-covariance.  This observation is important for planning data analysis efforts.}


\maketitle
%

\section{Introduction}

Solar acoustic waves are randomly excited by turbulent convection in the upper convection zone. They propagate through the interior and are refracted by the increase of sound speed with depth. By measuring the travel times of wave packets between pairs of points on the solar surface we can probe the subsurface structure and dynamics of the Sun.  

The travel time between any two points on the solar surface is measured from the cross-covariance of the oscillation signals observed at these two points. The definition of the temporal cross-covariance function between points $\vec{x_1}$ 
and $\vec{x_2}$ is 
\begin{eqnarray}
C(\vec{x_1}, \vec{x_2}, \tau) = \frac{h_t}{T-|\tau|} \sum_{t} 
\phi^{*} (\vec{x_1},t) \phi (\vec{x_2},t+\tau) \; ,\label{eq:ccdef}
\end{eqnarray}
where $\phi (\vec{x}_i, t)$ is the oscillation signal at time $t$ and  position $\vec{x}_i$ on the solar surface,  $T$ is the duration of the observations, $h_t$ is the time sampling, and $\tau$ is the time lag. 
\cite{1993Natur.362..430D} demonstrated that wave travel times can be used to map flows and sound-speed heterogeneities in the solar interior
\citep[for a review see][]{2005LRSP....2....6G}.
The cross-covariance function has, however, intrinsic noise due to the 
stochastic nature of solar oscillations. Understanding the statistical properties of this noise is crucial for interpreting measurements of wave travel times \citep{2004ApJ...614..472G,2014A&A...567A.137F}.

In practice, wave travel times are estimated by fitting a model to 
the cross-covariance function.
In the standard approach, the fitting parameters $\vec{p}$ are obtained by  minimizing a merit function $X(\vec{p})$
using the least-squares method, where
\begin{eqnarray}
  X(\vec{p}) = \sum_i f(t_i) \left[C_{\mathrm{obs}}(t_i)-C_{\mathrm{model}}(t_i; \vec{p}) \right]^2  \ 
\label{eq:chisq_constc}\end{eqnarray}
and $f(t_i)$ is the window function
(e.g., to isolate the first bounce part of the cross-covariance function).
The function  $C_{\mathrm{obs}}(t_i)$ is the observed cross-covariance 
function at time lag $t_i$ and
$C_{\mathrm{model}}(t_i; \vec{p})$ is the model 
cross-covariance function. 
\cite{1997SoPh..170...63D} and \cite{1997ASSL..225..241K} used
a Gabor wavelet with five parameters to model the cross-covariance function, 
while \cite{2002ApJ...571..966G} performed one-parameter fits to measure the time shift compared to a reference cross-covariance function.
\cite{2004ApJ...614..472G} simplified the definition of \cite{2002ApJ...571..966G} in the limit of small travel-time shifts; this linearized form is more robust to noise.

As an alternative procedure, \cite{2006ApJ...640..516C} proposed using
\begin{equation}
  X(\vec{p}) = \sum_i \frac{f(t_i)}{\sigma_i^2} \left[C_{\mathrm{obs}}(t_i)-C_{\mathrm{model}}(t_i; \vec{p}) \right]^2 \; , \label{eq:chisq_sigma}
  \end{equation}
where $\sigma_i^2$ is the variance of the cross-covariance function at time lag $t_i$.  \cite{2006ApJ...640..516C}  estimated the variance using a Monte-Carlo approach and the method proposed by \citet{2004ApJ...614..472G}:  many realizations of the cross-covariance function were generated using a model that depends only on the observed oscillation power spectrum.   The assumptions of the model are temporal stationary and spatial homogeneity. 
\cite{2006ApJ...640..516C} mentioned that Eqs.~\ref{eq:chisq_constc} and \ref{eq:chisq_sigma}
lead to different travel times in the short-distance case, but did not explain why, nor did they discuss the dependence of $\sigma_i$ on time lag $t_i$.

In this paper we discuss some properties of the noise in the cross-covariance function. First, we evaluate the variance and covariance of the cross-covariance function in a stochastic oscillation model.  We then show that for some standard  situations (far field) the variance of the cross-covariance function is nearly independent of both time lag and distance between the observation points, while this is not the case in the near field.  

We note that in this paper we discuss the
two-point (point-to-point) cross-covariance function.
In typical helioseismology travel-time measurements, 
point-to-annulus cross-covariance functions are 
widely used to increase the signal-to-noise ratio; 
the oscillation signal averaged over an annulus and 
the oscillation signal at the central point of the annulus 
are used to calculate the cross-covariance function. 
Discussions to such geometry are found in \cite{2004ApJ...614..472G}.


\section{Variance of the cross-covariance function}

In this paper we use a stochastic oscillation model to determine the variance and covariance of the cross-covariance function.  We use the model of \cite{2004ApJ...614..472G}, which is itself a generalization of
the standard model for realization noise in global helioseismology 
\citep[e.g.][]{1984PhDT........34W, 2000MNRAS.319..365A}.  
Throughout this paper we use the notation of \cite{2004ApJ...614..472G}.

The observed signal $\phi (\vec{x}, t)$  is sampled with temporal cadence $h_t$ and  spatial sampling $h_x$ over a  duration $T = N_t h_t$ and an area $L^2$ with  $L = N_x h_x$.
Using the horizontal wavevector $\vec{k}$ and the angular frequency $\omega$,
we denote the Fourier transform of the oscillation signal by 
\begin{eqnarray}
\phi(\vec{k}, \omega) = \frac{{h_x}^2 h_t}{(2\pi)^3} \sum_{\vec{x},t} \phi (\vec{x}, t )
e^{-i(\vec{k}\cdot \vec{x} -\omega t)}  \label{eq:FTdef1}
\end{eqnarray}
and the inverse Fourier transform by
\begin{eqnarray}
\phi(\vec{x}, t) = {h_k}^2 h_{\omega} \sum_{\vec{k},\omega} \phi (\vec{k}, \omega )
e^{i(\vec{k}\cdot \vec{x} -\omega t)} \; ,\label{eq:FTdef2}
\end{eqnarray}
where $h_k = 2 \pi/L$ and $h_{\omega} = 2\pi/T$.
The notation $\sum_{t}$ means that the sum is over times $t_j = j h_t$ where $j$ is an integer
in the range $[-N_t /2, N_t/2-1]$. 
Here we assume $N_t$ is even. 
Note that since $\phi (\vec{x}, t )$ is real, we have  $\phi(-\vec{k},- \omega) =\phi^*(\vec{k}, \omega)$. 
Using the Fourier transform of $\phi$, we can rewrite the cross-covariance function 
(Eq.~\ref{eq:ccdef}) as
\begin{eqnarray}
C(\vec{x_1}, \vec{x_2}, \tau) =  ({h_k}^2 h_{\omega})^2
\sum_{\vec{k},\vec{k}^{\prime}, \omega}
\phi^{*} (\vec{k},\omega) \phi (\vec{k}^{\prime},\omega) 
e^{-i (\vec{k}\cdot \vec{x_1} - \vec{k}^{\prime} \cdot \vec{x_2})}e^{-i\omega \tau} \; .
\end{eqnarray}
For the sake of simplicity we assumed $\phi$ is cyclic in $t$, i.e., $\phi(t+T) = \phi(t)$. A generalization is possible by applying zero-padding.

Let us define $\mathcal{P} (\vec{k},\omega)$ as the expectation value (ensemble average) of the power spectrum,
\begin{eqnarray}
\mathcal{P} (\vec{k},\omega) = E [| \phi (\vec{k},\omega) |^2] \; ,
\end{eqnarray}
where $E[X]$ denotes the expectation value of $X$.
This quantity has the symmetry $\mathcal{P} (-\vec{k},-\omega) =\mathcal{P} (\vec{k},\omega) $ as the oscillation signal is real valued.  Assuming every mode $(\vec{k},\omega)$ is excited stochastically and independently,  we model the Fourier transform of the oscillation signal as 
\begin{eqnarray}
\phi (\vec{k},\omega) = \sqrt{\mathcal{P}(\vec{k},\omega)} \mathcal{N} (\vec{k},\omega) ,
\end{eqnarray}
where $\mathcal{N}(\vec{k},\omega)$ is a centered complex Gaussian random variable with unit variance and independent real and imaginary parts, i.e., $E[ \mathcal{N}(\vec{k},\omega)] =0$, 
$E[\mathcal{N}^*(\vec{k},\omega) \mathcal{N}(\vec{k},\omega)] =1$,
and $E [  \mathcal{N}^* (\vec{k},\omega) \mathcal{N} (\vec{k}^{\prime},\omega^{\prime}) ] =0$ if $\vec{k} \ne \vec{k}^{\prime}$ or $\omega \ne \omega^{\prime}$, but with the additional requirement that $\mathcal{N} (-\vec{k},- \omega) =\mathcal{N}^* (\vec{k}, \omega)$ to comply with the condition that $\phi (\vec{x}, t)$ is real.
In this case, $|\phi (\vec{k}, \omega)|^2$ has a chi-square distribution with two degrees of freedom. This model is rather general. In frequency space, it is known observationally  to be a good description of solar oscillations \citep{1984PhDT........34W,1998A&AS..132..107A,2000MNRAS.319..365A}. 

In this simple model (spatial homogeneity) 
the distribution function of the cross-covariance
function is a normal distribution \citep{2010Nagashima}.
As a consequence, the statistical properties of the cross-covariance function are determined completely by its expectation value and variance. The expectation value of the cross-covariance function, $\mathcal{C}$, is proportional to the inverse Fourier Transform of the power spectrum:
\begin{equation}
\mathcal{C}(\vec{x_2} - \vec{x_1} , \tau) \equiv E[C(\vec{x_1},\vec{x_2},\tau)] =  ({h_k}^2 h_{\omega})^2
\sum_{\vec{k}, \omega} \mathcal{P} (\vec{k},\omega) 
e^{i \{\vec{k}\cdot (\vec{x_2} - \vec{x_1})-\omega \tau\}} 
= ({h_k}^2 h_{\omega})\mathcal{P}(\vec{x_2} - \vec{x_1} , \tau).
\label{eq:EC}
\end{equation}
The covariance of the cross-covariance 
function is given by
\begin{eqnarray}
\begin{aligned}
&\mathrm{Cov} [ C (\vec{x_1},\vec{x_2},\tau)  ,C(\vec{x_1}^{\prime}, \vec{x_2}^{\prime}, \tau^{\prime})] =   ({h_k}^2 {h_\omega})^4
\sum_{\vec{k}, \vec{k}^{\prime}, \omega} 
\mathcal{P}(\vec{k},\omega) \mathcal{P}(\vec{k^{\prime}},\omega) 
\Bigl[
e^{ -i \{\vec{k}\cdot(\vec{x_1}-\vec{x_1}^{\prime} ) - \vec{k}^{\prime}
\cdot (\vec{x_2} - \vec{x_2}^{\prime}) + \omega (\tau-\tau^{\prime})\}}+e^{ -i\{ \vec{k}\cdot(\vec{x_1}-\vec{x_2}^{\prime} ) + \vec{k}^{\prime}
\cdot (\vec{x_1}^{\prime} - \vec{x_2})+  \omega (\tau+\tau^{\prime}) \} }
\Bigr] . \label{eq:covC}
\end{aligned}
\end{eqnarray}
Here the covariance of two complex variables $X$ and $Y$ is
defined by $\mathrm{Cov} [X,Y] \equiv E [XY^*]-E[X]E[Y^*]$. 
Detailed derivations of Eq.~\ref{eq:covC} are shown by \cite{2010Nagashima} and \cite{2014A&A...567A.137F}.
For the case when 
$\vec{x_1} =\vec{x_1}^{\prime}$, $\vec{x_2}=\vec{x_2}^{\prime} \equiv \vec{x_1}+\vec{\Delta} $, and $\tau=\tau^{\prime}$,
this simplifies  to the variance of the cross-covariance function:
\begin{eqnarray}
\mathrm{Var} [ C(\vec{x_1},\vec{x_2},\tau)]&=&   ({h_k}^2 {h_{\omega}})^4  \sum_{\vec{k},\vec{k}^{\prime},\omega} 
\mathcal{P} (\vec{k},\omega) \mathcal{P} (\vec{k^{\prime}},\omega)
\left[ 1 + e^{ 
i\{ (\vec{k}+\vec{k}^{\prime}) \cdot \vec{\Delta} -
\omega (2\tau)\} } \right] \nonumber \\
&=&  
\mathcal{Q} (\vec{0}, 0 )+ \mathcal{Q}(\vec{\Delta}, \tau ) , \label{eq:CCvariance}
\end{eqnarray}
where 
$\mathrm{Var}[X] \equiv \mathrm{Cov}[X,X]$ and 
we have defined 
\begin{eqnarray}
\mathcal{Q}(\vec{\Delta}, \tau ) \equiv {h_k}^4 {h_\omega}^3
\sum_{\omega} h_\omega e^{-2i\omega\tau} \left( \sum_{\vec{k}} h_k^2 
\mathcal{P} (\vec{k},\omega) e^{i \vec{k}\cdot\vec{\Delta}}\right)^2 \; .\label{eq:Qdef}
\end{eqnarray}
The noise of the cross-covariance function is given by 
$\sigma (\vec{\Delta},\tau)  =
\sqrt{\mathcal{Q} (\vec{0}, 0 )+ \mathcal{Q}(\vec{\Delta}, \tau )}$. 
Notice that in Eq.~\ref{eq:CCvariance} only the second term, $\mathcal{Q}(\vec{\Delta}, \tau )$,
depends on $\vec{\Delta}$ and $\tau$.
This term  $\mathcal{Q}(\vec{\Delta}, \tau)$ can be written in terms of
$\mathcal{C}(\vec{\Delta}, \tau)$ defined by Eq.~\ref{eq:EC}: 
\begin{eqnarray}
\mathcal{Q}(\vec{\Delta}, \tau )=\frac{1}{N_t} \sum_{\tau'} \mathcal{C}(\vec{\Delta},\tau')
\mathcal{C}(\vec{\Delta}, 2\tau-\tau') .  \label{eq:Q_CC}
\end{eqnarray}
This is consistent with the  
discussions in \cite{2004ApJ...614..472G},
except that the terms involving $\mathcal{F}$ of Eq.~C8 of 
\cite{2004ApJ...614..472G} should vanish from the exact solution.

In the next section we compute $\mathcal{Q}(\vec{\Delta}, \tau )$ for some standard situations in helioseismology analyses. 
Note that  $|\mathcal{Q}(\vec{\Delta}, \tau )| \ll \mathcal{Q}(\vec{0}, 0)$ would imply that $\sigma$  is  independent of  time lag and the distance.

\section{Examples} 
\label{sec:example}

In this section, we consider two examples to
show the behavior of $\mathcal{Q}(\vec{\Delta}, \tau )$.
First, we look at oscillations with a simple Gaussian distribution of power in $k$-$\omega$ space.
Second, we consider  oscillations with a solar-like  power distribution.  Also, by using Eq.~\ref{eq:Q_CC}, we provide a simple formula for $\mathcal{Q}$ in terms of $\mathcal{C}$. 

\subsection{Wave power localized in wavenumber-frequency space} \label{sec:example_gauss}

First, we consider the simplest power distribution: power localized around $(k_x,k_y,\omega) = \pm (k_0, 0,\omega_0)$.
Even in this simplest case, however, $\mathcal{Q}(\Delta, \tau )$ cannot be computed analytically. We simplify the computation of Eq.~\ref{eq:Qdef} by approximating the discrete sums 
over $k$ and $\omega$ with continuous integrals.  
This approximation is good when 1) the wavenumber resolution ($h_k \propto 1/N_t$) and frequency resolution ($h_\omega \propto 1/N_t$) are small enough to capture the 
smallest scales in the power spectrum and 2) the spatial and temporal sampling rates ($h_x$ and $h_t$) are small enough so that the resulting Nyquist frequencies in space and time are larger than $k_0$ and $\omega_0$.

Consider a Gaussian power distribution of the form 
\begin{equation}
\mathcal{P}(\vec{k},\omega) 
= \left\{
\begin{array}{cl} 
\frac{1}{2\pi \sigma_k \sigma_\omega} 
\left(
e^{-(k_x-k_0)^2/(2\sigma_{k}^2)}  
e^{-(\omega-\omega_0)^2/(2\sigma_{\omega}^2) } 
+  
e^{-(k_x+k_0)^2/(2\sigma_{k}^2)}  
e^{-(\omega+\omega_0)^2/(2\sigma_{\omega}^2) }
\right) \ , & k_y=0 \\
0 \ , & k_y \neq 0
\end{array}
\right.
\label{eq:Power_1D_1peak}
\end{equation}
The second peak of power is required to ensure that  the oscillation signal is real.
With this power distribution, 
\begin{equation}
\mathcal{Q}(\Delta,\tau)
= \frac{ {h_k}^4 e^{-\sigma_k^2\Delta^2-\sigma_\omega^2 \tau^2}}{\sqrt{\pi}\sigma_\omega}  
\left( e^{-\omega_0^2/\sigma_\omega^2}  +\cos[2(\omega_0 \tau-k_0 \Delta)] \right)  ,
\label{eq:inf_cont_Q_1Dgauss1}
\end{equation}
where $\Delta$ here is the distance in the $x$ direction. 
This means that the ratio of the two terms that make up the variance of the cross-covariance function is given by
\begin{equation}
\frac{\mathcal{Q}(\Delta, \tau)}{\mathcal{Q}(0,0)}=
e^{ -\sigma_k^2\Delta^2-\sigma_\omega^2\tau^2 }
\frac{e^{-\omega_0^2/\sigma_{\omega}^2}  +\cos{2(\omega_0 \tau-k_0 \Delta)}}{e^{-\omega_0^2/\sigma_{\omega}^2}  +1} .
\label{eq:inf_cont_Q_1Dgauss1_ratio}
\end{equation}
Equation \ref{eq:inf_cont_Q_1Dgauss1_ratio} shows that
the variance of the cross-covariance function has a peak around the origin with a width of $(1/\sigma_k, 1/\sigma_\omega)$ in space and time, equal to the coherence scales of the oscillations. If $\Delta$ and $\tau$ are large compared to $1/\sigma_k$ and $1/\sigma_\omega$,  then 
the variance of the cross-covariance function (Eq.~\ref{eq:CCvariance}) is independent of both $\Delta$ and $\tau$.
Therefore, the remaining task is to discuss  the coherence length and coherence time.

\begin{figure}[thp]
\centering
\includegraphics{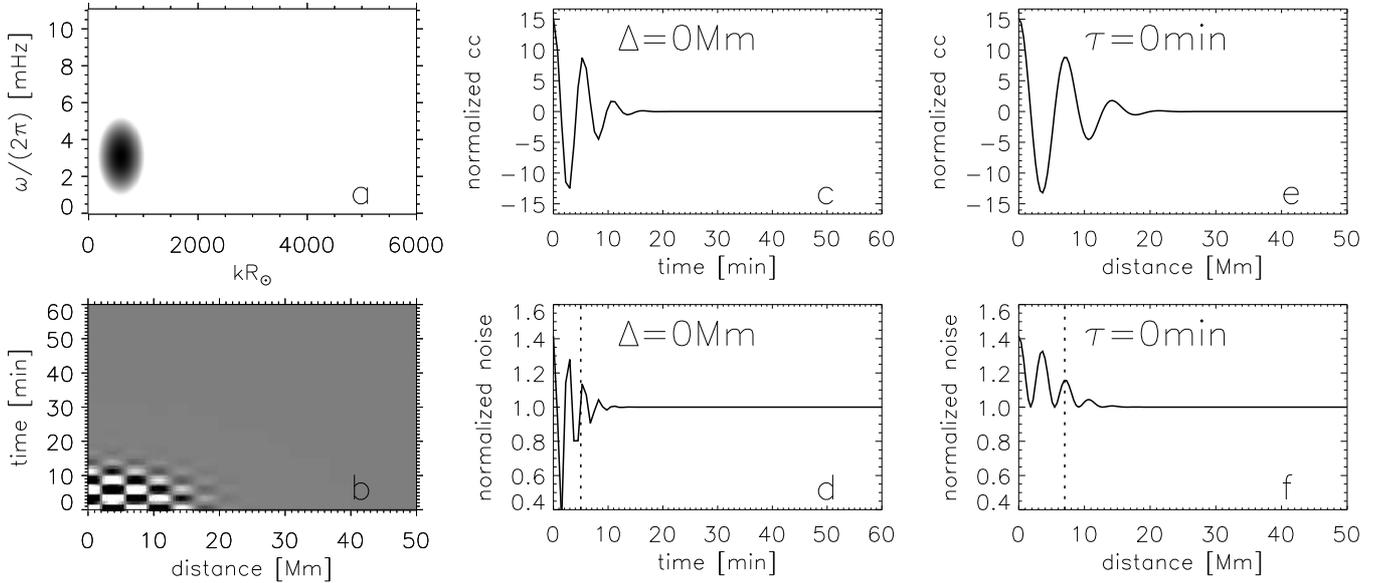}
\caption{Examples of the power spectrum in logarithmic 
gray scale (Panel a) and the 
expectation value of the cross-covariance function (Panel b) in the case of the 
Gaussian power spectrum. In Panel a (and the same is true in  Figs.~\ref{fig:Example_Obs2d_1}a and \ref{fig:Example_Obs2dVph_1}a), larger and smaller power is indicated by black and white, respectively. 
The peak of the power is located at $(k_0 R_\odot,\omega_0/(2\pi)) = (600, 3 \ \mathrm{ mHz})$ and 
the peak has widths $(\sigma_k R_\odot, \sigma_{\omega}/(2\pi))= (100, 0.5 \ \mathrm{ mHz})$. 
 The cuts through the expectation value of the cross-covariance function
 at $\Delta=0$ and $\tau=0$ are shown in Panels~c and~e,
and the noise for the same cuts are shown in Panels~d and~f.
The cross-covariance (cc) function and the noise, $\sigma (\Delta,\tau)=\sqrt{\mathcal{Q}(0,0)+\mathcal{Q}(\Delta, \tau)}$, are both
normalized by $\sqrt{\mathcal{Q}(0,0)}$. This choice of normalization means that
the amplitude of the cross-covariance function directly gives the signal-to-noise ratio ($E[C(\Delta,\tau)]/\sigma(0,0)$).
The vertical dotted lines in Panels d and f indicate the expected width of the noise in time and space, 5~minutes and 7~Mm, respectively.}
\label{fig:ExamplePureGauss_1}
\end{figure}

Figure~\ref{fig:ExamplePureGauss_1} shows an
example with a Gaussian power distribution in 2D. 
This example has a single Gaussian-shaped peak at
 ($k_0 R_\odot,\omega_0/(2\pi)) = (600, 3 \ \mathrm{mHz})$ with the widths of $(\sigma_{k} R_\odot, \sigma_{\omega}/(2 \pi))= (100,0.5 \ \mathrm{mHz})$, where $R_\odot$ is the solar radius. We chose these parameters as typical values of the Sun. 
Note that in Fig.~\ref{fig:ExamplePureGauss_1}a we show only $k \ge 0$ and $\omega \ge 0$, but there is also the associated peak at $(-k_0, -\omega_0)$.
In this case the widths in space and time 
($1/\sigma_{k}$ and $1/\sigma_{\omega}$ from Eq.~\ref{eq:inf_cont_Q_1Dgauss1_ratio})
are 7~Mm and 5~minutes, respectively
(see vertical dotted lines in
Figs.~\ref{fig:ExamplePureGauss_1}d and \ref{fig:ExamplePureGauss_1}f).
For the temporal and spatial scales larger than these coherence scales (a standard case in helioseismology analysis),
the noise level is constant, i.e. $|\mathcal{Q}(\Delta,\tau)| \ll \mathcal{Q}(0,0)$.

Note that this simple Gaussian power distribution with 
typical solar values 
also explains the origin of the `horizontal stripes'  that are seen in the near field (Fig.~\ref{fig:Example_Obs2d_1}b) in the time-distance diagram.

Thus a Gaussian is a simple but useful model for the envelope of the power spectrum, which determines the coherence scales in time and space. The solar oscillation power spectrum has many peaks, however; we can calculate the detailed behavior of the cross-covariance function  using Eq.~\ref{eq:inf_cont_Q_1Dgauss1} and model power spectra with multiple Gaussian peaks as well; more detailed investigation is found in Appendix \ref{sec:multipeak}.

\subsection{Solar-like oscillation power spectrum}

\begin{figure}[pt]
\centering
\includegraphics{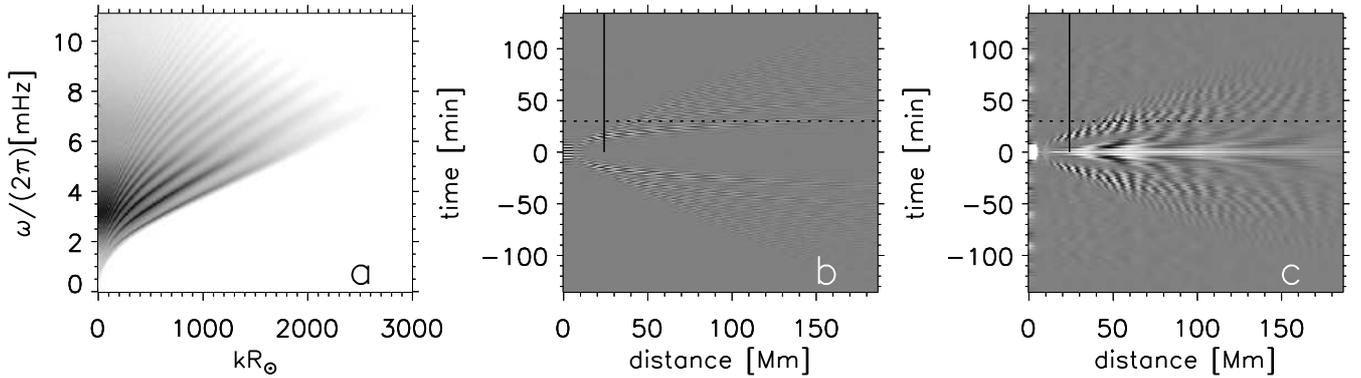}
\caption{
Example of the case of p-mode power spectrum of HMI Doppler observations: power spectra (Panel a) in 
logarithmic gray scale, 
cross-covariance function $\mathcal{C}(\Delta,\tau)$ (Panel b), and $\mathcal{Q}(\Delta,\tau)$ (Panel c). 
The cuts at the solid vertical and dashed horizontal lines on the Panels b are shown
in Fig.~\ref{fig:Example_Obs2d_2}.
}
\label{fig:Example_Obs2d_1}
\end{figure}

\begin{figure}[pt]
\centering
\includegraphics{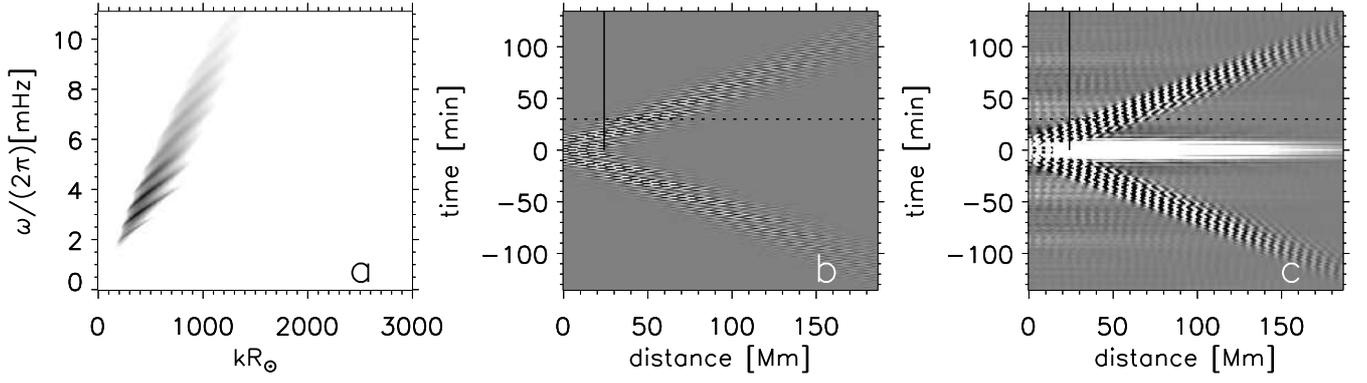}
\caption{Similar to Fig. \ref{fig:Example_Obs2d_1} but for the case of 
p-mode power spectrum of HMI Doppler observation datacube with a phase speed filter centered at $v_{\mathrm{ph}}=36 $ km s$^{-1}$ with the width of 5 km s$^{-1}$. The central phase speed corresponds to a ray which has 2-degree 
(24.1-Mm) skip distance on the Sun. 
}
              \label{fig:Example_Obs2dVph_1}
    \end{figure}

Here we consider power spectra of solar oscillations observed by the Helioseismic and Magnetic Imager \citep[HMI;][]{2012SoPh..275..229S}
on the Solar Dynamics Observatory. 
In one case we consider the full p-mode power spectrum.
In another case  we apply a phase-speed filter to isolate waves with a    skip distance on the Sun of $2^\circ$ ($24.1$ Mm) \citep{2010Nagashima}: the filter is centered at the horizontal phase speed of $v_{\mathrm{ph}}=36$ km s$^{-1}$ and has a width of 5~km s$^{-1}$.  For these two cases, the power spectra and the cross-covariance functions are shown  in Figs.~\ref{fig:Example_Obs2d_1} and \ref{fig:Example_Obs2dVph_1}. 
To construct the power spectra, we used observations obtained from 18 UT on January 22, 2011 to 12 UT on January 26, 2011. We divided this data set into ten nine-hour segments, and in each segment we tracked the quiet region near disk center at the Carrington rotation rate using the code mtrack \citep{2011JPhCS.271a2008B}.
The mean of the azimuthally-averaged power spectra is an estimate of the expectation value  $\mathcal{P}(k,\omega)$. 
The pixel scale is $0.03$ heliographic degrees ($0.36$ Mm),  the temporal sampling cadence is 45 s, and the field of view is 1024 pixels square.
Before calculating the power spectra, we take the running difference in time  for detrending, and apply spatial and temporal  zero-padding to handle the non-cyclic functions in our formula.

\begin{figure}[t]
\centering
\includegraphics{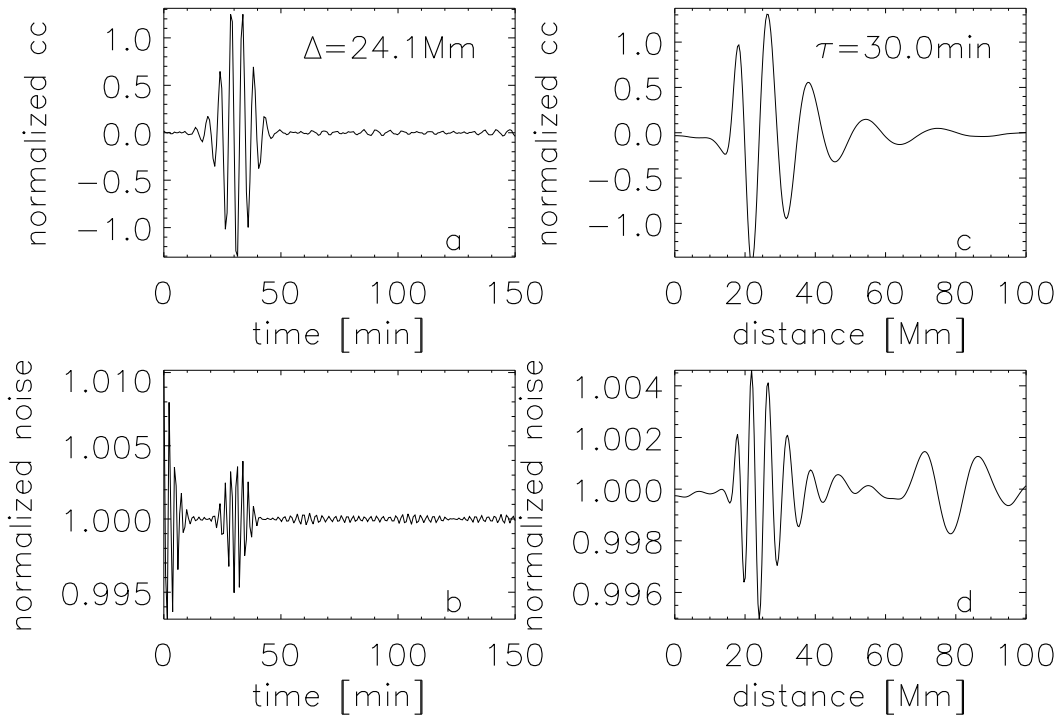}
 \caption{
The expectation value of the cross-covariance function 
 (Panels a,c) and its noise, $\sigma (\Delta,\tau)$, (Panels b,d) 
 for the full p-mode power spectrum of HMI Doppler observations 
(Fig. \ref{fig:Example_Obs2d_1}).
Panels a and b are cuts at $\Delta=24.1$ Mm,  
and Panels c and d are cuts at $\tau=30.0$ minutes.
The cross-covariance (cc) function and the noise are both
normalized by $\sqrt{\mathcal{Q}(0,0)}$. 
}
\label{fig:Example_Obs2d_2}
 \end{figure}

\begin{figure}[ht]
\centering
\includegraphics{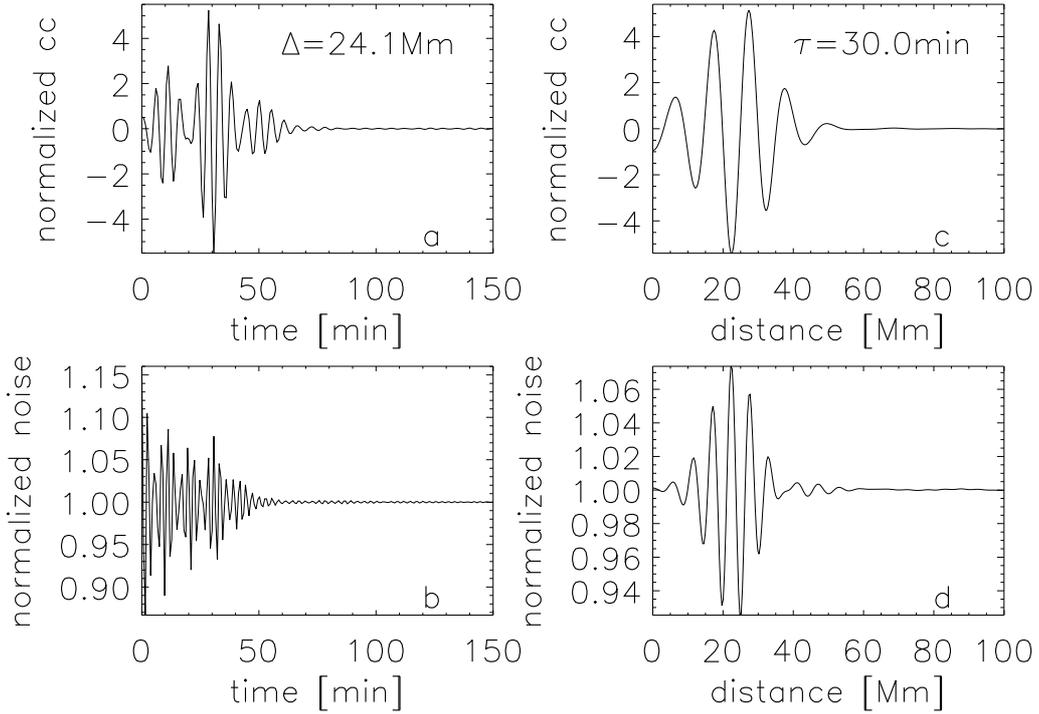}
\caption{
The expectation value of the cross-covariance function (Panels a,c) and 
 its noise (Panels b,d) and for the power spectrum of HMI Doppler observation 
 datacube with a phase speed filter (Fig.~\ref{fig:Example_Obs2dVph_1}).
Panels a and b are cuts at $\Delta=24.1$ Mm,  
and Panels c and d are cuts at $\tau=30.0$ minutes.
Normalization factors are determined in the same way as Fig.~\ref{fig:Example_Obs2d_2}.
}
 \label{fig:Example_Obs2dVph_2}
    \end{figure}

Figs.~\ref{fig:Example_Obs2d_2} and \ref{fig:Example_Obs2dVph_2} show
cuts through the cross-covariance functions and the noise in cross-covariance function at $\Delta=24.1$~Mm and $\tau= 30$~min.
In these cases the non-constant part of the noise is small;
for example, in Fig.~\ref{fig:Example_Obs2d_2}b 
variations in noise are only $0.4$\% of the constant part of the noise.
These examples show that the noise in the cross-covariance is independent of time lag (and thus that, in the far field, Eq.~\ref{eq:chisq_sigma} can be reduced to Eq.~\ref{eq:chisq_constc}).
As we mentioned in Sec.~\ref{sec:example_gauss}, 
the coherence scale of the solar oscillations
is about 7 Mm and 5 minutes. In local helioseismology we usually care only about scales larger than these coherence scales. 

By comparing Figs.~\ref{fig:Example_Obs2d_2}b and \ref{fig:Example_Obs2dVph_2}b,
or Figs.~\ref{fig:Example_Obs2d_2}d and \ref{fig:Example_Obs2dVph_2}d,
it is evident that the noise in the cross-covariance function is reduced by the phase-speed filter. 
Here we define a signal-to-noise ratio for the cross-covariance  by
$\mathcal{C}(\Delta,\tau) = E[C(\Delta,\tau)]$ is the signal and $\sqrt{\mathcal{Q}(0,0)}$ is 
the noise level. 
Since throughout this paper we choose the normalization factor as $\sqrt{\mathcal{Q}(0,0)}$, 
the amplitude of the cross-covariance function in the figures directly gives the 
signal-to-noise ratio. 
The signal-to-noise ratio is $1.3$ in the case without the phase-speed filter (Fig.~\ref{fig:Example_Obs2d_2}a), while it is more than $5$ in the case with the phase-speed filter (Fig.~\ref{fig:Example_Obs2dVph_2}a) .
But with a phase-speed filter, the noise variations
extend to larger time lags and the amplitude of the variations is larger as well.
Therefore, we need to carefully choose the filter,
considering the trade-off between the signal-to-noise ratio and the 
variation of the noise with time-lag and distance.
This is consistent with what \cite{2013SoPh..287...71D} reported about the width of the phase-speed filter and the 
signal-to-noise ratio of the travel time.

\subsection{Noise estimate from the cross-covariance function}

In time-distance helioseismology 
we measure travel times from the cross-covariance function.
Therefore, it would be practical if we could estimate the noise not from the power spectrum 
but instead from the cross-covariance function itself.
Here we show a simple example.

Since Eq.~\ref{eq:Q_CC} tells us that the noise function is 
written with a simple form using the expectation value of the 
cross-covariance function, 
in order to estimate the noise level 
of the cross-covariance function, namely $\sqrt{Q(0,0)}$, in practice, 
we need to obtain the zero-distance cross-covariance function
and fit it to obtain the parameters to estimate $\mathcal{Q}(0,\tau)$
in addition to calculation of the cross-covariance functions at targeted distances.
The cross-covariance function is often approximated by a Gabor wavelet
\citep[e.g.,][]{1997SoPh..170...63D,1997ASSL..225..241K}.
Therefore, if we fit the zero-distance cross-covariance function with 
\begin{equation}
C(0,\tau) = A_0 e^{-\tau^2/(2 {\sigma_{g,0}^2})} \cos(\omega_0 \tau)  , 
\end{equation}
then we obtain using Eq.~\ref{eq:Q_CC}\footnote{Here we approximate the discrete sum 
as an integral (from $-\infty$ to $\infty$) over a continuous time.}
\begin{equation}
\mathcal{Q}(0,\tau) = \frac{A_0^2 \sigma_{g,0}}{4\sqrt{\pi}}  e^{-\tau^2/(2{\sigma_g}^2)}
\{ e^{-{\omega_0}^2 \sigma_{g,0}^2} +\cos(2\omega_0 \tau) \}  .
\end{equation}
Once the fitting parameters $A_0$, $\sigma_{g,0}$ and $\omega_0$ are obtained, 
the noise level is estimated as 
\begin{equation}
 \sqrt{\mathcal{Q}(0,0)}= \frac{A_0 \sqrt{\sigma_{g,0}}}{2\sqrt[4]{\pi}}  \sqrt{ e^{-{\omega_0}^2 \sigma_{g,0}^2} +1}. \label{eq:Noisesimplewavelet}
\end{equation}

Moreover, when we use the cross-covariance function 
at a certain targeted distance $\Delta_c$, 
we can also estimate the oscillations of the noise around $\sqrt{\mathcal{}Q(0,0)}$. 
In that case, if we consider a pair of Gabor wavelets as a 
symmetrical expectation value of the cross-covariance function
at a certain distance $\Delta_c$ :
\begin{equation}
\mathcal{C}(\Delta_c, \tau) = A \left( e^{-(\tau-\tau_g)^2/(2 {\sigma_g}^2)} \cos{\omega (\tau-\tau_p)} 
+  e^{-(\tau+\tau_g^2/(2 {\sigma_g}^2)} \cos{\omega (\tau+\tau_p}  )\right) \ , \label{eq:Csimplewavelet}
\end{equation}
where $A$,$\tau_g$, $\sigma_g$, $\tau_p$, $\sigma_p$ are fitting parameters at $\Delta=\Delta_c$, 
according to Eq.~\ref{eq:Q_CC}
\begin{equation}
\begin{aligned}
\mathcal{Q}(\Delta_c, \tau) = \frac{A^2 \sigma_g}{4\sqrt{\pi}} 
&\biggl[ e^{- (\tau-{\tau_g})^2/\sigma_g^2} 
\left\{\cos{2\omega(\tau-\tau_p)} + e^{-\omega^2 {\sigma_g}^2}  \right\} 
+ e^{- (\tau+{\tau_g})^2/\sigma_g^2} 
\left\{\cos{2\omega(\tau+\tau_p)} + e^{-\omega^2 {\sigma_g}^2}  \right\}   \\
&  + 2 e^{-\tau^2/{\sigma_g}^2} \left\{\cos{(2\omega\tau)} + e^{-\omega^2 {\sigma_g}^2} \cos{2\omega(\tau_p -\tau_g ) } \right\}
\biggr]. 
\end{aligned} \label{eq:Qsimplewavelet}
\end{equation}
Hence, the oscillatory part of the variance basically consists of 
Gaussians peaking at $\tau=\tau_g$, $-\tau_g$, and $0$.
The width of these peaks is determined by the wavepacket width ($\sigma_g$),
and the frequency of the oscillatory part of the noise is twice higher
than that of the cross-covariance.
This is consistent with the calculation results
we showed in the previous subsection.
The oscillation field with a narrower peak in the power distribution 
has a cross-covariance function with the broader wavelet form, 
and thus, the width of the noise is broader as well.

\section{Conclusions and outlook}

In this work, by modeling stochastic solar oscillations
we calculated the variance of the point-to-point cross-covariance function 
as a function of time-lag and distance between the two observation points.  
As a result, we showed that the variance of the cross-covariance in the far-field
is independent of both time-lag and distance. 
We also showed in the previous section that the constant noise level can be estimated 
using the fitting parameters of the zero-distance cross-covariance function.
The fact that in the far field the noise is flat means that the signal-to-noise ratio 
for the cross-covariance function is proportional to the amplitude of 
the expectation value of the cross-covariance, and this is of importance in analysis.

As mentioned in the introduction the full statistics of the cross-covariance function 
\citep[e.g.,][]{2004ApJ...614..472G, 2012SoPh..276...19J, 2014A&A...567A.137F}
are  needed to optimize the definition of travel time.  
In particular, it would be appropriate to obtain the parameters of the model, $\vec{p}$, by minimizing the merit function
\begin{eqnarray}
X (\vec{p}) =  \left\| \Lambda^{-1/2} f(t) [C(t)-C_{\mathrm{model}} (t; \vec{p})] \right\|^2, \label{eq:chisq_general}
\end{eqnarray}
where  $C(t)$ is the cross-covariance function at time lag $t$,
$C_{\mathrm{model}}(t; \vec{p})$ is the model cross-covariance function, and
\begin{eqnarray}
\Lambda_{ij} = \mathrm{Cov}[C(t_i), C(t_j)] 
\end{eqnarray}
is the covariance matrix of the cross-covariance function.
Note that the merit function, Eq.~\ref{eq:chisq_general}, is the  general form of Eq.~\ref{eq:chisq_sigma}.
The computation of $\Lambda$ is future work. In the inversion process we cannot avoid the computation of $\Lambda$, and the covariance of the travel time. For those extended calculations we perhaps could use the concept of our calculation to simplify the computation of the variance of the covariance in this paper and \cite{2014A&A...567A.137F}. An alternative and simpler approach would be transform to the Fourier domain, where different frequencies are uncorrelated.

\begin{acknowledgements}
We thank Damien Fournier and Jesper Schou for useful discussions. 
Part of this work was done while K.N. was supported by
the Research Fellowship from the Japan Society for the Promotion of Science 
for Young Scientists.
K.N. and L.G. acknowledge support
from EU FP7 Collaborative Project “Exploitation of Space Data for Innovative Helio- and
Asteroseismology” (SPACEINN).
The HMI data used are courtesy of NASA/SDO and the HMI science team. 
The German Data Center for SDO (GDC-SDO), funded by the German Aerospace Center (DLR),
provided the IT infrastructure to process the data.
\end{acknowledgements}

\bibliographystyle{aa} 

\appendix

\section{Analytical calculation of the noise for a multi-peak power distribution} \label{sec:multipeak}
If the power distribution has not only one pair of peaks but multiple pairs of peaks,
and if the number of the Gaussian peaks is $N_p$, then the power is 
\begin{eqnarray}
\mathcal{P}(\vec{k},\omega) 
=\sum_{l=0}^{N_p-1} A_l \Biggl\{ \frac{1}{\sqrt{2\pi} \sigma_{k,l}} e^{-(k_x-k_{l})^2/(2\sigma_{k,l}^2)}
\frac{1}{\sqrt{2\pi} \sigma_{\omega,l}} e^{-(\omega-\omega_l)^2/(2\sigma_{\omega,l}^2) } +  \frac{1}{\sqrt{2\pi} \sigma_{k,l}} e^{-(k_x+k_l)^2/(2\sigma_{k,l}^2)}
\frac{1}{\sqrt{2\pi} \sigma_{\omega,l}} e^{-(\omega+\omega_l)^2/(2\sigma_{\omega,l}^2) } \Biggr\} \ 
\end{eqnarray}
at $k_y=0$ and zero at $k_y \neq 0$., 
where $A_l \ (l=0, 1, \dots N_p-1)$ are real-valued amplitudes.

In this case, with a straightforward calculation\begin{eqnarray}
\mathcal{Q}(\Delta,\tau) 
&=& \sum_{l=0}^{N_p-1} {h_k}^4 {A_l}^2 \frac{e^{- \sigma_{k,l}^2 \Delta^2} e^{-\sigma_{\omega,l}^2 \tau^2}}{\sqrt{\pi}\sigma_{\omega,l}}  
\{e^{-\omega_l^2/\sigma_{\omega,l}^2}  +\cos{2(\omega_l \tau-k_l \Delta) } \} \nonumber \\
&&+ 2\sum_{l=0}^{N_p-2}  \sum_{m=l+1}^{N_p-1} {h_k}^4 
\frac{A_l A_m e^{-{(\sigma_{k,l}}^2 +{\sigma_{k,m}}^2) \Delta^2/2}}
{\sqrt{2\pi(\sigma_{\omega,l}^2+\sigma_{\omega,m}^2)}} e^{-2\sigma_{\omega,l}^2 \sigma_{\omega,m}^2 \tau^2/(\sigma_{\omega,l}^2+\sigma_{\omega,m}^2)}\times \nonumber \\
&& \Biggl\{
e^{-(\omega_m-\omega_l)^2/(2(\sigma_{\omega,l}^2+\sigma_{\omega,m}^2))} \cos{ \left\{(k_l+k_m)\Delta + 2\tau (\sigma_{\omega,m}^2\omega_l+\sigma_{\omega,l}^2\omega_m)/
(\sigma_{\omega,l}^2+\sigma_{\omega,m}^2) \right\} } \nonumber \\
&&+ e^{-(\omega_m+\omega_l)^2/(2(\sigma_{\omega,l}^2+\sigma_{\omega,m}^2))} \cos{\left\{(k_l-k_m)\Delta + 2\tau (\sigma_{\omega,m}^2\omega_l-\sigma_{\omega,l}^2\omega_m)/
(\sigma_{\omega,l}^2+\sigma_{\omega,m}^2) \right\} }  \Biggr\} \label{eq:multipeak_Q}
\end{eqnarray}
where $\Delta$ is in the $x$ direction, and 
\begin{eqnarray}
\mathcal{Q}(0,0) =\sum_{l=0}^{N_p-1} {h_k}^4 {A_l}^2 \frac{e^{-\omega_l^2/\sigma_{\omega,l}^2} +1}{\sqrt{\pi}\sigma_{\omega,l}}  + 2\sum_{l=0}^{N_p-2} \sum_{m=l+1}^{N_p-1} {h_k}^4
\frac{A_l A_m }
{\sqrt{2\pi(\sigma_{\omega,l}^2+\sigma_{\omega,m}^2)}} 
 \Biggl\{
e^{-(\omega_m-\omega_l)^2/(2(\sigma_{\omega,l}^2+\sigma_{\omega,m}^2))}  + e^{-(\omega_m+\omega_l)^2/(2(\sigma_{\omega,l}^2+\sigma_{\omega,m}^2))}  \Biggr\} . 
\end{eqnarray}

Although the form of $\mathcal{Q}(\Delta, \tau)$ (Eq.~\ref{eq:multipeak_Q}) is not that simple,
all the terms in the summation have a Gaussian envelope 
centered at the origin in space and time. 
Since the oscillatory cosine functions have respective spatial and temporal frequencies, 
in the case of  sufficiently many modes (large $N_p$), the sum, and thus 
$\mathcal{Q}(\Delta, \tau)/\mathcal{Q}(0,0)$ will damp more rapidly  
than each mode component, except near the origin.

\end{document}